\documentclass[aps,prb,twocolumn,showpacs,groupedaddress]{revtex4}

\usepackage{graphicx}
\usepackage{dcolumn}
\usepackage{bm}

\begin{document}

\preprint{PRB/Biex1}

\title{Multiexcitons confined within a sub-excitonic volume:
Spectroscopic and dynamical signatures of neutral and charged biexcitons 
in ultrasmall semiconductor nanocrystals}

\author{M. Achermann}
\author{J.~A. Hollingsworth}
\author{V.~I. Klimov}
\affiliation{Chemistry Division, C-PCS, MS-J585, Los Alamos National 
Laboratory, Los Alamos, NM 87545, USA}%


\begin{abstract}
The use of ultrafast gating techniques allows us to resolve both 
spectrally and temporally the emission from short-lived neutral and negatively charged 
biexcitons in ultrasmall (sub-10 nm) CdSe nanocrystals 
(nanocrystal quantum dots). 
Because of ``forced" overlap of electronic wave functions and reduced 
dielectric screening, these states are characterized by giant interaction 
energies of tens (neutral biexcitons) to hundreds (charged biexcitons) of meV. 
Both types of biexcitons show extremely short lifetimes (from sub-100 picoseconds 
to sub-picosecond time scales) that rapidly shorten with decreasing nanocrystal size.  
These ultrafast relaxation dynamics are explained in terms of highly 
efficient nonradiative Auger recombination.

\end{abstract}


\maketitle

In ultrasmall, sub-10 nm semiconductor nanocrystals 
[known also as nanocrystal quantum dots (QDs)], new 
types of strongly interacting multiexciton states can exist that do not occur 
in ``natural" bulk semiconductors. The interaction strength in the 
multiexciton system scales directly with the excitation density 
(i.e., the number of excitons per cm$^3$). However, in bulk materials, 
there is a fundamental limit on maximum densities for which excitons 
still exist. This limit corresponds to the exciton dissociation 
threshold ($n_{th}$) which is roughly determined by densities 
corresponding to one exciton per excitonic volume: $n_{th} \sim a^{-3}_B$, 
where $a_B$ is the exciton Bohr radius.  Above this threshold, a 
dielectric excitonic gas transforms into metallic 
electron-hole (e-h) plasmas, and the Coulomb interactions become greatly 
reduced because of strong dynamic screening by unbound charge 
carriers.\cite{mott}

In contrast to the bulk case, in 
semiconductor nanoparticles one can generate states in which several 
excitons occupy a volume comparable to or smaller than the volume of 
a bulk exciton. Such ``squeezed" exciton states are characterized by 
greatly enhanced multiparticle interactions resulting from a {\it 
forced} overlap of electronic wavefunctions and reduced 
dielectric screening. The latter effect occurs because of a 
penetration of the electric field outside a nanoparticle (the 
surrounding medium is typically characterized by a lower dielectric 
constant) and reduced efficiency of dynamic screening  (carriers are 
locked in the nanoparticle by rigid boundary conditions which 
result in low electronic polarizability).

In addition to revealing new physics, studies of strongly 
interacting multiexcitons are relevant to several emerging 
technologies. For example, optical 
amplification and lasing in sub-10 nm nanocrystals relies on 
emission from particles occupied with two or 
more excitons.\cite{Klimov1, KlimovAPL} Therefore, the control of optical gain 
properties of nanocrystal QDs requires detailed 
understanding of energy spectra and dynamics of strongly confined 
multiexcitons. Another example involves quantum technologies in which 
two excitons in a quantum dot can be 
used as a quantum-bit pair\cite{Steel1} for quantum 
information processing or as a source of entangled 
photon pairs.\cite{Yamamoto1}

The challenge in experimentally detecting spectroscopic signatures of 
strongly confined multiexcitons is associated with their very short 
(picoseconds to hundreds of picoseconds) lifetimes that are limited by 
nonradiative, multiparticle Auger recombination. \cite{Klimov2} 
Because these times are significantly shorter than the radiative 
decay time, multiexcitons  are undetectable in
time-integrated (cw) photoluminescence  (PL) spectra. In this article, 
we report for the first time the emission spectra of strongly 
confined neutral and charged biexcitons in sub-10 nm CdSe QDs detected 
using time-resolved, femtosecond PL measurements. These multiexciton 
states are manifested in femtosecond PL spectra as two bands 
observed in addition to a single-exciton emission line. Both bands 
are characterized by a superlinear pump-intensity dependence 
indicating their multiexcitonic origin. The detected states are 
also characterized by extremely large interaction energies (tens to 
hundreds of meV) which can be explained by the existence of strong 
{\it attractive} forces between squeezed excitons. Since the 
observed interaction energies are comparable to or even greater than
room-temperature thermal energies, ultrasmall, sub-10 nm quantum 
dots can be used for
room-temperature implementations of quantum technologies that are 
based on multiexciton interactions.

QDs used for this study are highly monodisperse, chemically 
synthesized CdSe nanocrystals overcoated with a ZnS shell and a final 
layer of organic ligand molecules (core-shell nanocrystal QDs). We 
investigated 7 samples with a narrow size dispersion of 5--7\% and 
mean dot radii from 1.1 to 3.6 nm which corresponds to 0.2--0.8 
$a_B$. The samples are excited at 
400 nm by the frequency-doubled output of an amplified Ti:sapphire 
laser (250 kHz repetition rate). Ultrafast, time-resolve PL 
measurements are performed using a femtosecond PL up-conversion (uPL) 
technique.\cite{shah} In these measurements, the emission from QDs 
is frequency-mixed (gated) with 200 fs pulses of the fundamental 
laser radiation in a nonlinear-optical $\beta$-barium borate crystal. 
The sum frequency signal is spectrally filtered with a monochromator 
and detected using a cooled photomultiplier tube coupled to a photon 
counting system. The time resolution of uPL measurements 
is $\sim$300 fs. All measurements are performed at room temperature.

The need for a short time-resolution for detecting multiexciton 
states in strongly confined QDs becomes evident from data in 
Fig. 1(a), in which we compare a
cw PL spectrum with the uPL spectra measured at 
$\Delta t=1$ ps and 200 ps after excitation. All these spectra were 
recorded at the same pump fluence ($w_{p}$) of 3.4 mJcm$^{-2}$, which corresponds 
to the excitation of more than 10 excitons per dot on average. 
Because of fast nonradiative Auger recombination, all multiexcitons 
decay on a sub-100 ps time scale\cite{Klimov2} and, therefore, the 
uPL spectrum at $\Delta t=200$ ps [circles in Fig. 1(a)] is entirely 
due to single exciton emission. Interestingly, this spectrum is 
essentially identical to the cw spectrum [shaded area in Fig. 1(a)], 
indicating that time-integrated emission is dominated by single 
excitons even in the case of high excitation levels for which several 
excitons are initially generated in a significant number of dots.
The early-time uPL spectrum recorded at $\Delta t=1$ ps [solid line 
in Fig. 1(a)] is distinctly different from the excitonic emission and 
displays a clear shoulder on the low-energy part of the excitonic 
band and a new high-energy emission band. These new features only develop 
at high excitation densities that correspond to an average 
number of e-h pairs (excitons) per dot  
$\overline{N}>1$   
and, therefore, should be attributed to multiexciton states as analyzed
below [$\overline{N}$ is estimated 
from the expression 
$\overline{N} = \sigma_a(w_p/\hbar\omega_p)$, where $\sigma_a$ is 
the QD absorption cross section at the 
pump spectral energy $\hbar \omega_p$; Ref. 8]. 

\begin{figure}[h] 
\includegraphics[width=8cm]{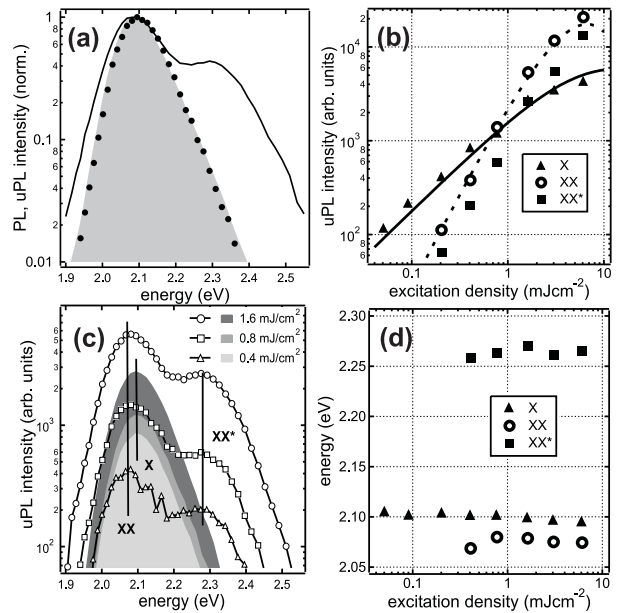}
\caption{
(a) Normalized time-integrated (shaded area) and time-resolved uPL spectra
($w_{p}=3.4$ mJcm$^{-2}$) measured at
$\Delta t=1$ ps (solid line) and $\Delta t =200$ ps (solid circles). 
(b) Pump-intensity dependence of X (triangle), XX (circles), and 
XX* (squares) band amplitudes compared to fitting curves (lines) calculated 
assuming the Poisson distribution of QD populations (see text for details). 
(c) Single exciton (shaded areas) and multiexciton  (symbols) 
emission spectra (extracted from the 1 ps uPL spectra) 
at different excitation densities. (d) Pump-intensity dependence of the 
spectral positions of the X (triangle), XX (circles)
and  XX* (squares) bands. 
}
\end{figure}

In the analysis of high-pump-intensity PL spectra, we assume 
the Poisson distribution of initial QD populations. This distribution 
accurately describes photoexcited QD ensembles 
in the case of femtosecond pumping well above 
the energy gap when the probability for generating a new e-h 
pair in a dot is independent of the number of pairs 
already existing in this dot.\cite{KlimovJPC} For the
Poisson distribution, the concentration of QDs that contain 
$N$ excitons ($n_N$) immediately following excitation 
with an ultrashort laser pulse is expressed as
$n_N=n\overline{N}^N e^{-\overline{N}}/{N!}$, 
where $n$ is 
the total concentration of QDs in the sample. 
While the intrinsic 
decay of singly excited CdSe nanocrystals occurs due to radiative
recombination, which is characterized by
the time constant $\tau_r \approx$ 20 ns at 
room temperature,\cite{crooker:2003:1} 
the multiexciton 
decay is primarily due to much faster nonradiative Auger recombination with 
time constants $\tau_A \lesssim$ 200 ps for the QD sizes 
studied.\cite{Klimov2} 
Since the decay of 
QD multiexcitons eventually produces a singly excited dot, 
the concentration of excitons ($n_X$) at intermediate time 
delays  $\tau_A< \Delta t < \tau_r$ is determined 
by  the total number of dots excited by a pump pulse 
(independent on the initial number of excitations per dot) 
and, hence,
can be estimated from the expression
$n_X = n-n_0 = n(1-e^{-\overline {N}})$.
This expression closely reproduces the pump 
dependence of the PL intensity measured at the center of 
the exciton band at 
$\Delta t = $ 200 ps [compare solid triangles (experiment) and 
a solid line (model) in Fig. 1(b)] indicating the validity of the 
above considerations. 

In order to extract a multiparticle component from uPL spectra, 
one needs to account for a contribution from singly-excited QDs 
at early times after excitation. We calculate this contribution by 
scaling the purely excitonic spectra detected at 
time delays $\Delta t > \tau_A$ 
($\Delta t$ = 200 ps in our experiments) by two factors. One 
factor accounts for 
a single exciton decay (derived from the PL dynamics 
measured at low excitation intensities corresponding to  
$\overline{N}<1$) and the other one describes a contribution from 
additional 
single exciton states produced as a result of the multiexciton decay. 
In the case of the Poisson distribution,
the latter factor can be calculated as   
$n_1/n_X = \overline{N}/(e^{\overline{N}}-1)$.

In Fig. 1(c), we show, by shaded areas, pump dependent uPL spectra recorded at 
$\Delta t$ = 200 ps; as discussed earlier, these spectra 
correspond to the purely single exciton emission. 
By scaling the 200-ps spectra according to the procedure 
described above and subtracting them from early time spectra measured at 1 ps, 
we extract the spectral component that is entirely due to the multiexciton emission 
[symbols in Fig. 1(c)]. The multiexciton spectra show two bands: one (XX) on the low and 
the other (XX*) on the high energy side of the exciton (X) line 
(the multiexciton band notations are clarified below). 
The positions of XX and XX* bands\cite{note2} are independent of the 
pump power [Fig. 1(d)], indicating that each of these bands is due to 
emission from dots in a well-defined multiparticle state rather than 
due to superposition of emission spectra from dots with different 
numbers of photoexcitations (in the latter case, the positions/shapes 
of the spectra should show pump dependence).

In Fig. 1(b), we compare the pump dependence of the XX and XX* bands 
(open circles and solid squres, respectively) with that 
measured for a single exciton band (solid triangles). 
While the single exciton emission shows 
a linear initial growth (with respect to $w_p$), 
the initial growth is {\it quadratic} for both multiexciton bands, 
indicating an excitation mechanism that involves absorption of two 
photons from the same pump pulse. This quadratic mechanism is also confirmed 
by the fact that the
pump dependence over the entire range of intensities  
(including the saturation region at high pump levels) 
can be closely modeled in terms of the Poisson 
distribution assuming that the mutliexciton emission is due to doubly excited QDs 
[i.e., is described by the  term 
$n_2 = ({n \overline{N}}^2/2)e^{-\overline{N}}]$. 

Despite identical pump dependencies, the XX and XX* bands 
likely originate from different type of biexciton states 
because of their very different spectral positions.
The low energy XX band 
is located immediately below  the single-exciton line, which allows 
us to assign it to the emission of a biexciton, in which both 
electrons are in the lowest 1S state (the 1S$^2_e$ biexciton). Such 
biexcitons are generated via absorption of two photons by a dot that 
was not occupied prior to the arrival of a pump pulse. The biexcitonic assignment 
of the XX feature is also confirmed by the analysis of its dynamics. As shown
below, size dependent decay times of the XX band closely match those measured for 
two electron-hole pair states 
in Ref. 6 
using a transient absorption experiment.  

The spectral position of the second, high-energy multiexciton band 
indicates that it likely involves the emission from the excited (1P) 
electron state. Because of fast, sub-picosecond 
1P-to-1S relaxation,\cite{KlimovPRL} the occupation of the 1P state can only be 
stabilized if the 1S orbital is fully filled (i.e., contains two 
electrons). This suggests that the high-energy band is not due to excited neutral 
biexcitons but rather due to charged biexciton states (XX*) involving two holes and three electrons, with 
the electron-shell configuration 1S$^2_e$1P$^1_e$. The observed XX* emission originates likely from 
the 1P$_e$-1S$_h$ transition, which is nominally forbidden, but can be allowed in multiexciton 
states due to Coulomb interactions.\cite{KochJOSA}

The fact that 
the XX* states are excited via a ``quadratic" process can be 
explained by the existence of a
sub-ensemble of dots with long-lived 1S electrons that ``survive" on 
time scales comparable to or longer than the time separation between 
two sequential pump pulses (4 $\mu$s in our case). Since excitons 
decay radiatively on a nanosecond time scale, long-lived 1S electrons 
are likely associated with charge-separated 
electron-hole pairs formed, e.g., as a result of hole surface 
trapping.\cite{KlimovJPC} The efficiency of surface trapping 
correlates with PL quantum yields (QYs), suggesting that the intensity 
of the XX* band should increase as the QY is decreased. This is exactly 
the trend we observe in our experiments using samples with differently 
prepared surfaces (e.g., purely organic passivation vs. ZnS overcoating).  
Furthermore, the existence of dots that contain 
long-lived charges (charged dots) has been previously suggested based 
on results of single-QD PL intermittence\cite{Efros2} and QD 
charging experiments.\cite{Brus}  

Our model that explains two multiexciton features in terms of 
two sub-ensembles of {\it neutral} and {\it charged} dots is 
schematically depictured in Fig. 2. In this model, both the excitonic 
(X) and the biexcitonic (XX) features originate from neutral dots 
(no charges prior to the arrival of a pump pulse) while the 
charged biexciton (XX*) feature is generated by adding two excitons into a 
charged dot that already contains an electron in the 1S state and 
a hole in a surface state.

\begin{figure}[h] 
\includegraphics[width=8cm]{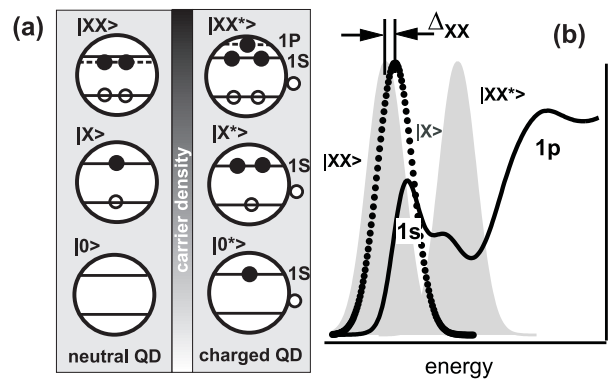}
\caption{(a) A cartoon illustrating a progression of states that 
develop in neutral and charged dots with increasing pump level. (b) A 
schematic illustration of neutral (XX) and charged (XX*) biexciton 
emission spectra compared to a single exciton PL spectrum (X) 
(solid circles) and a typical absorption spectrum of CdSe 
nanocrystal QDs (1S and 1P are two strong optical transitions that 
involve the lowest and the first excited electron states 
 and $\Delta_{XX}$ is the biexciton binding energy).
}
\end{figure}

Since we clearly observe XX* species, one would expect that 
charged excitonic X* species should also be observed in our 
experiments. The fact that this type of states is not clearly 
manifested in uPL spectra is likely because their emission 
spectrally overlaps with the emission from either X or XX states. On 
the other hand, we do observe some indirect dynamical signatures 
of X* species such as a fast initial decay of the PL signal observed 
when nominally less than one exciton per dot is 
excited. In addition to ultrafast processes such as carrier surface 
trapping, this fast initial decay may be due to ultrafast Auger 
recombination of charged excitons.

In bulk CdSe, excitons are weakly attractive and form biexcitons 
with a binding energy $\delta E_2\approx 4.5$ meV. \cite{Shionoya} 
Strong 3D quantum confinement and reduced dielectric screening 
have been expected to strongly modify exciton-exciton interactions in QDs. 
However, the few available theoretical studies on this 
topic,\cite{Efros3,Zunger2,Takagahara,Banyai,Koch} 
have yielded conflicting results regarding both the sign (repulsion vs. attraction) 
and the magnitude of the exciton-exciton interaction energy, 
highlighting the need for experimental benchmarks. 
Early experimental attempts to address this problem relied on indirect transient 
absorption measurements applied to QD/glass composites with relatively poor QD size 
monodispersity ($>20\%$ size dispersion).\cite{peyghambarian,KlimovPRB} 
These experiments, along with more recent work on epitaxial 
CdSe QDs,\cite{Forchel} 
provided indications for a strong enhancement  in the biexciton binding energy in QDs compared to bulk CdSe.   

By applying our ultrafast PL experiments to a set of almost monodisperse QD 
samples spanning a wide range of sizes, we directly evaluate the magnitude 
and the size dependence of the exciton-exciton interaction energy. 
The shift of the biexciton emission band with respect to the single-exciton peak ($\Delta_{XX}$) is determined by the sum 
of the exciton-exciton interaction energy ($\delta E_2$) and 
the relaxation energy of the exciton that is created as a result of biexciton decay. 
If we assume that the biexcitons preferentially decay into ground state, 
relaxed excitons (as suggested, e.g., by calculations in Ref. 17), 
the ``biexcitonic" shift $\Delta_{XX}$  provides a direct measure of the biexciton binding energy.

The QD  size-dependent biexcitonic shift (presumably equivalent to $\delta E_2$) is plotted in Fig. 3(a). 
For the largest dots studied in this work ($R=35\AA$ ), $\Delta_{XX} = 14$  
meV, which is close to the bulk exciton binding energy and several 
times greater than the binding energy of a bulk biexciton. As the dot 
size is decreased, the binding energy first increases up to  33 meV at 
$r =18$ $\AA$, then it starts to decrease and is approximately 12 meV for $r =11$ $\AA$. 
The initial increase of $\Delta_{XX}$ follows the $1/r$ dependence 
(dashed line) as expected for exciton-exciton Coulomb interactions. The 
opposite trend observed at very small sizes likely results from 
repulsive
electon-electron and hole-hole interactions that overwhelm the 
exciton-exciton attraction in the regime of extremely strong spatial 
confinement. \cite{Takagahara}

\begin{figure}[h]  
\includegraphics[width=8cm]{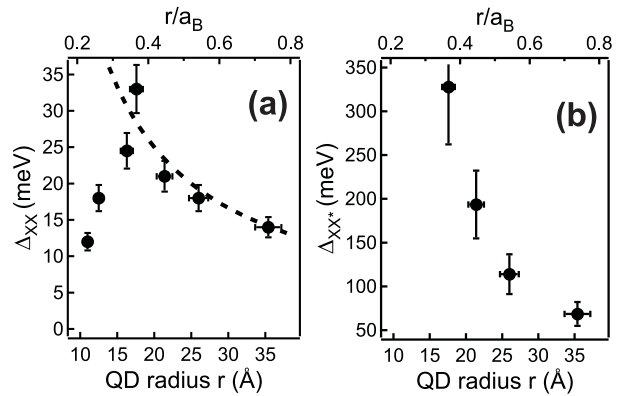}
\caption{QD size dependence of the neutral $\Delta_{XX}$ (a) and 
charged $\Delta_{XX^{*}}$ (b) biexciton energy shifts $(a_{B}\simeq48$ $\AA$ in bulk CdSe). 
The $\Delta_{XX}$ size dependence [panel (a)] in the range $r \ge 17\AA$ 
is fit to $1/r$ (dashed line). 
}
\end{figure}

To estimate the interaction energy of a charged biexciton state, 
we find the difference between energy of the XX* band and the energy 
of the 1P single exciton.  Because of the extremely fast 1P-to-1S 
relaxation in the single exciton regime,\cite{KlimovPRL} the latter 
quantity cannot be measured experimentally.  Therefore, we estimate 
it as a sum of the measured 1S exciton energy (the center of the uPL 
band at long time after excitation) and the energy spacing between 
the 1S and 1P electronic states from Ref. 24. 
 For sizes from 35 to 18 $\AA$, for which we 
were able to resolve the XX* feature, the interaction energy $\Delta_{XX^{*}}$ of the 
XX* state monotonously increases from 70 to 330 meV.   Extremely large 
magnitudes of $\Delta_{XX^{*}}$  likely result from large densities of 
uncompensated positive and negative charges within the XX* species. 
These large charge densities are generated because of strongly 
different spatial distributions of electron (involving both 1S and 1P 
states) and hole (involving 1S and surface states) wave functions.

The uPL experiment allows us also to directly measure lifetimes of 
multiexciton states (Fig. 4). Because of highly efficient Auger 
recombination, the decay of neutral and charged biexcitons is very 
fast and occurs on sub-100 ps time scales [Fig. 4(a)]. The decay time 
constant for both neutral and charged biexcitons exhibits the $R^3$ 
dependence on the QD size [Fig. 4(b)], consistent with results 
of previous transient absorption studies.\cite{Klimov2} 
As the dot size decreases from 35 to 11 \AA~
the neutral biexciton life time shortens 
from 100 to 6 ps; and these time constants are close to those 
reported in Ref. 6 
for two-pair states. 
The time constant for charged biexcitons varies from 20 to 0.7 ps for 
the same range of sizes. Interestingly, the Auger decay of charged 
biexcitons occurs much faster than the decay of 
neutral biexcitons. This difference likely results 
from stronger exciton-exciton interactions characteristic of charged 
species.

\begin{figure}[h]  
\includegraphics[width=8cm]{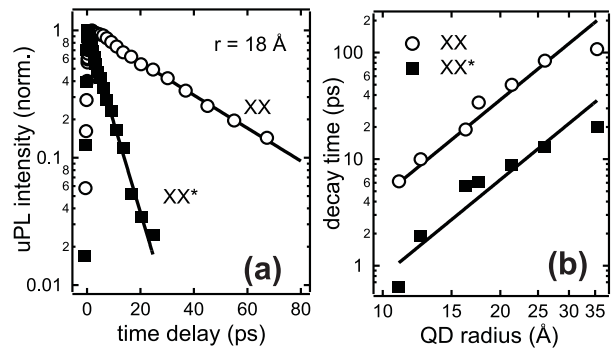}
\caption{(a) Neutral and charged biexciton dynamics for QDs with $r=18\AA$ 
after subtracting a  slowly decaying background due
to single exciton decay.
(b) Neutral (circles) and charged (squares) biexciton life times as a 
function of QD radii fit to the $r^{3}$ dependence.}
\end{figure}

The large interaction energies measured in this work provide an interesting opportunity for 
achieving lasing in ultrasmall, 
sub-10 nm QDs in the single exciton regime. In the absence of Coulomb 
interactions, the optical gain in these QDs develops at 
excitation densities $\overline{N} >1$, implying 
that light amplification is provided by doubly excited nanoparticles 
(i.e., biexcitons). However, if exciton-exciton interactions are present in 
the system and the biexcitonic shift is greater than the transition
linewidth, the gain threshold can be achieved for $\overline{N} >2/3$, i.e., before the 
onset of biexciton generation. If the above conditions are met 
(through, e.g., improved sample monodispersity and, possibly, 
engineered biexcitonic interactions), one could solve a major problem in 
the field of nanocrystal QD lasing, namely the ultrafast gain decay 
due to multiparticle Auger recombination.

In conclusion, by applying ultrafast, PL up-conversion technique, 
we are able to detect the emission from short-lived neutral and charged 
biexcitons in ultrasmall CdSe nanocrystals with radii 
from 1.1 to 3.6 nm (0.2-0.8 $a{_B}$). The analysis of the spectral 
characteristics of the emission indicates that both types of biexcitons 
are characterized by extremely large interaction energies 
that are on the order of tens and hundreds of meV, for neutral and 
charged species, respectively. These values are much greater than the 
biexciton binding energy in bulk CdSe (4.5 meV) indicating a significant 
increase in Coulomb exciton-exciton interactions induced by strong quantum confinement. 
The biexciton states show extremely fast relaxation dynamics on a 100-ps time scale. 
The observed decay constants scale proportionally to the nanocrystal volume and 
are in the range from 6 to 100 ps for neutral biexcitons and from 0.7 to 20 ps 
for charged biexciton. These ultrafast relaxation dynamics are due to 
highly efficient Auger recombination resulting from strong interparticle Coulomb interactions.

This work was supported by Los Alamos Directed Research and
Development Funds, and the U. S. Department of Energy, Office of Sciences,
Division of Chemical Sciences.

\pagebreak

\pagebreak


\begin{thebibliography}{99}
\bibitem{note2}
The line widths of the multiexciton bands are comparable to the width 
of the single-exciton emission spectrum, suggesting that there is no size 
selection within the quasi-monodisperse ensemble during the generation of multiexcitons. 
Therefore, the ensemble mean multiexciton emission energies can be 
characterized by the centers of the XX and XX* bands.
\end{thebibliography}
\end{document}